\documentclass[
 aip,
 amsmath,amssymb,
 reprint,%
 nofootinbib
]{revtex4-1}

\usepackage{graphicx}
\usepackage{dcolumn}
\usepackage{bm}
\usepackage{enumerate}
\usepackage{siunitx}
\usepackage{amsmath}
\usepackage{cases}
\usepackage{xcolor}
\usepackage{slashed}

\DeclareSIUnit{\calorie}{cal}

\newcommand\beq{\begin{equation}}
\newcommand\eeq{\end{equation}}
\renewcommand{\d}{\mathrm{d}}

\newcommand{\angstrom}{\text{\normalfont \AA}}

\newcommand{\PR}[1]{{\color{black} #1}}

\newcommand{\tin}{\textrm{in}}

\newcommand{\f}{\textrm{f}}
\newcommand{\p}{\textrm{p}}
\newcommand{\out}{\textrm{out}}
\newcommand{\D}{\textrm{D}}
\newcommand{\ex}{\textrm{ex}}
\newcommand{\kB}{k_\textrm{B}}

\graphicspath{{figures/}}

\begin{document}


\title{Ion filling of a one-dimensional nanofluidic channel in the interaction confinement regime}

\author{Paul Robin}%
\author{Adrien Delahais}
\author{Lyd\'eric Bocquet}
\affiliation{Laboratoire de Physique de l'\'Ecole Normale Sup\'erieure, ENS, Universit\'e PSL, CNRS, Sorbonne Universit\'e, Universit\'e Paris Cit\'e, Paris, France}
\author{Nikita Kavokine}
\email{nikita.kavokine@mpip-mainz.mpg.de}
\affiliation{Department of Molecular Spectroscopy, Max Planck Institute for Polymer Research, Ackermannweg 10, 55128 Mainz, Germany}
\affiliation{Center for Computational Quantum Physics, Flatiron Institute, 162 5$\rm ^{th}$ Avenue, New York, NY 10010, USA}

\date{\today}

\begin{abstract}
Ion transport measurements are widely used as an indirect probe for various properties of confined electrolytes. It is generally assumed that the ion concentration in a nanoscale channel is equal to the ion concentration in the macroscopic reservoirs it connects to, with deviations arising only in the presence of surface charges on the channel walls. Here, we show that this assumption may break down even in a neutral channel, due to electrostatic correlations between the ions arising in the regime of interaction confinement, where Coulomb interactions are reinforced due to the presence of the channel walls. We focus on a one-dimensional channel geometry, where an exact evaluation of the electrolyte's partition function is possible with a transfer operator approach. Our exact solution reveals that in nanometre-scale channels, the ion concentration is generally lower than in the reservoirs, and depends continuously on the bulk salt concentration, in contrast to conventional mean-field theory that predicts an abrupt filling transition. We develop a modified mean-field theory taking into account the presence of ion pairs that agrees quantitatively with the exact solution \PR{and provides predictions for experimentally-relevant observables such as the ionic conductivity.} Our results will guide the interpretation of \PR{nanoscale ion transport measurements}.
\end{abstract}

\maketitle

\section{Introduction}

A channel connects two reservoirs filled with a salt solution at concentration $c_\out$. What is the salt concentration $c_\tin$ inside the channel? The straightforward answer $c_\tin = c_\out$ is challenged as soon as the channel's dimensions are at the nanometre scale~\cite{Schoch2008}. A deviation typically occurs because of the presence of a surface charge density $\Sigma$ on the channel walls. Indeed, a sufficiently long channel must remain electrically neutral~\cite{Levy2020}, which results in an imbalance of the concentrations $c_\tin^{\pm}$ of the positive and negative ions. In a cylindrical channel of radius $R$ that is smaller than the electrolyte's Debye length, the concentrations are given by the famous Donnan equilibrium result~\cite{Kavokine2021}: 
\begin{equation}
c_\tin^{\pm} = \sqrt{c_\out^2 + (2 \Sigma / R)^2} \pm 2\Sigma / R. 
\label{donnan}
\end{equation}
Eq.~\eqref{donnan} is widely used to infer a channel's surface charge from measurements of its conductivity at different salt concentrations. For sufficiently small surface charges ($2\Sigma/R \ll c_\out$), Eq.~\eqref{donnan} predicts $c_\tin = c_\out$ even at extreme nanoscales. Importantly, this prediction underlies the method for extracting confined ion mobilities from transport measurements, which has been applied down to 7-$\angstrom$-wide two-dimensional channels~\cite{Esfandiar2017}. Yet, physically, $c_\tin = c_\out$ stems from the assumption that the electrolyte solutions, both in the reservoirs and in the channel, behave as ideal gases of non-interacting ions. While such a description is valid in the bulk reservoirs at reasonable salt concentrations~\cite{Avni2022}, it must be challenged in the nanometre-scale channel which is subject to \emph{interaction confinement}~\cite{Kavokine2022} -- a reinforcement of the effective Coulomb interactions between the ions due to the dielectric contrast between the solvent (water) and the channel wall~\cite{Parsegian1969,Teber2005,Zhang2005,Levin2006,Kondrat2011,Loche2019,Kavokine2019,Robin2021,Kavokine2021,Kavokine2022}. 

\begin{figure*}
	\centering
	\includegraphics[width=\linewidth]{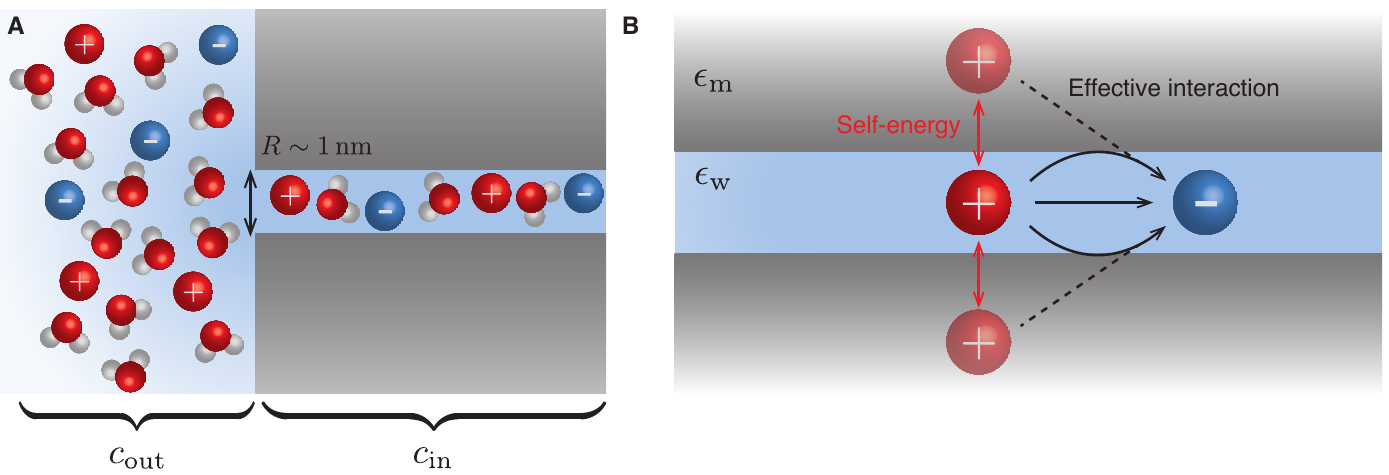}
	\caption{\textbf{Ion filling in the interaction confinement regime.} \textbf{A}. Schematic of the ion filling problem: a cylindrical nanochannel (radius $R \sim 1 \, \si{nm}$) is connected to macroscopic reservoirs of aqueous electrolyte. The salt concentration inside the channel, $c_\tin$, may differ from that in the reservoirs, $c_\out$. \textbf{B}. Physics of interaction confinement. When a charged species enters a nanochannel, the dielectric contrast between water ($\epsilon_{\rm w} \sim 80$) and walls ($\epsilon_{\rm m} \sim 2$) constraints the electric field lines to remain within the channel. This process can be interpreted in terms of image charges inside the channel walls, and results in an electrostatic self-energy barrier for ions to enter the channel, and reinforced interactions between ions.}
\end{figure*}

Due to interaction confinement, ions face a \emph{self-energy barrier} $E_{\rm s}$ when entering the channel~\cite{Parsegian1969,Teber2005}. It was first noted by Parsegian~\cite{Parsegian1969} that this should result in ion exclusion: \PR{the salt concentration within the channel is then given by an Arrhenius scaling }$c_\tin = c_\out e^{-E_{\rm s} / \kB  T}$ under the assumption of non-interacting ions. However, the result becomes more subtle as the confinement-reinforced ionic interactions are taken into account. Within a mean-field description of a spherical nanopore, Dresner~\cite{Dresner1974} predicted an abrupt filling transition, where $c_\tin$ was a discontinuous function of $c_\out$. Later, Palmeri and coworkers~\cite{Buyukdagli2010,Buyukdagli2010a} recovered a similar transition using a three-dimensional model of a cylindrical channel, treated within the variational field theory formalism of Netz and Orland~\cite{Netz2003}. While this approach could be applied to a realistic geometry, it took into account electrostatic correlations only approximately.

An exact treatment of electrostatic correlations is possible upon simplification of the geometry to a purely one-dimensional model, with the channel wall being taken into account by introducing an effective confined Coulomb interaction. The 1D electrolyte can then be mapped onto an Ising or 1D Coulomb-gas-type model; the transfer matrix solution of such models was used, for example, to discuss the capacitance of nanoporous systems~\cite{Demery2012,Lee2014,Demery2016a}. The lattice models may be taken to the continuum limit, and the resulting path integral solutions have been used to discuss various ion-exchange phase transitions that arise in the presence of fixed discrete charges inside the channel~\cite{Zhang2005,Zhang2006,Kamenev2006} and the ionic Coulomb blockade phenomenon~\cite{Kavokine2019}. Such models are particularly rich theoretically, as they support a mapping to non-Hermitian quantum mechanics~\cite{Gulden2021}. Nevertheless, to our knowledge, the fundamental problem of ion filling in an uncharged channel has not been tackled within this framework. 

In this paper, we treat the ion-filling problem in the interaction confinement regime using an exactly-solvable one-dimensional model. We find that the value of $c_\tin$ is strongly affected by the formation of Bjerrum pairs -- pairs of oppositely charged ions -- within the channel, which preclude the occurence of an abrupt filling transition. This is in contrast to the prediction of Palmeri and coworkers~\cite{Buyukdagli2010,Buyukdagli2010a}, and to the result of conventional mean-field theory. \PR{We then build on our exact results to propose a modified mean-field model that accounts for the relevant physical ingredients, and, particularly, for the presence of ion pairs.}

The paper is organized as follows. In Section II, we present the one-dimensional model and its solution within a path-integral formalism. The reader interested only in the physical outcomes may skip directly to Section III, where we discuss the model's prediction for the ion concentration within the channel, compare it to the mean-field solution, and interpret it in terms of tightly bound Bjerrum pairs. In Section IV, we establish a modified mean-field theory, based on the notion of \emph{phantom pairs}, that reproduces our exact solution. The mean-field theory allows us to determine the number of unpaired ions and produces experimentally relevant predictions for a nanochannel's ionic conductance. Section V establishes our conclusions. 

\section{1D Coulomb gas model}

\subsection{Confined interaction}

We consider a cylindrical channel of radius $R$ and length $L$, connected to macroscopic reservoirs (Fig.~1\textbf{A}). We first assume for simplicity that the channel is filled with water that has isotropic dielectric permittivity $\epsilon_{\rm w} = 80$, and that it is embedded in an insulating medium with much lower permittivity $\epsilon_{\rm m}$ (for a lipid membrane~\cite{Parsegian1969}, $\epsilon_m \sim 2$). The effective Coulomb interaction $V(x)$ between two monovalent ions separated by a distance $x$ on the channel axis can be computed exactly by solving Poisson's equation~\cite{Teber2005,Loche2019,Kavokine2019}. A simple approximate expression can be obtained for $x \sim R$ (ref.~\cite{Kavokine2021}): 
\begin{equation}
V(x) \approx \frac{e^2 \alpha}{2 \pi \epsilon_0 \epsilon_{\rm w} R} e^{-|x|/(\alpha R)}, 
\label{potential}
\end{equation}
where $\alpha$ is a numerical coefficient that depends on the ratio $\epsilon_{\rm w}/\epsilon_{\rm m}$ ($\alpha = 6.3$ for $\epsilon_{\rm w}/\epsilon_{\rm m} = 40$). \PR{The reinforcement of electrostatic interactions compared to the usual $e^2/4 \pi \epsilon_0 \epsilon_{\rm w}r$ Coulomb interaction that ions experience in bulk water can be interpreted in terms of images charges within the channel walls (Fig.~1\textbf{B}). Two confined ions interact not only with each other, but also with their respective image charges.}

We introduce the parameters $\xi \equiv \alpha R$ and $x_{ T} \equiv 2 \pi \epsilon_0 \epsilon_{\rm w} R^2 \kB  T / e^2$: both have the dimension of a length. With these notations, 
\begin{equation}
V(x) = \kB  T \frac{\xi}{x_{ T}} e^{-|x|/\xi}. 
\label{potential_xT}
\end{equation}
The effects of ion valence and of anisotropic dielectric response of confined water can be taken into account by adjusting $\xi$ and $x_{ T}$~\cite{Kavokine2019}. Formally, the expression in Eq.~\eqref{potential} is valid for any channel radius. Yet, it is only physically relevant if at $x\sim R$ the interaction is significant compared to $\kB  T$, which restricts in practice the applicability of Eq.~\eqref{potential} to $R \lesssim 2~\rm nm$. In such extreme 1D confinement, we may neglect the ions' degrees of freedom perpendicular to the channel axis and assume that they are constrained to move in one dimension. The partition function of such a 1D electrolyte may be computed exactly, as detailed in the next section. 

\subsection{Path integral formalism}

Here, we detail the analytical solution for the partition function of a 1D Coulomb gas-like system that was first introduced in ref.~\cite{Kavokine2019}. We set $\kB  T = 1$ until the end of Sec. II. We start from a lattice model, in order to rigorously establish a path integral description in the continuum limit. 

\begin{widetext}

Our computation is inspired by the original solution of the 1D Coulomb gas model by Lenard and Edwards~\cite{Edwards1962}, and subsequent studies by Demery, Dean and coworkers~\cite{Demery2016a,Demery2012,Demery2012a,Dean1997}, as well as Shklovskii and coworkers~\cite{Zhang2006,Kamenev2006}.   
We consider a one-dimensional lattice with sites $1,\dots,M$ as a model for the nanochannel of radius $R$ and length $L$. Each lattice site $i$ carries a spin $S_i$, which takes the values $\{0,1,-1\}$, corresponding respectively to no ion, a positive ion, or a negative ion occupying the site. We model the surface charge distribution as an extra fixed charge $q_i$ added at each lattice site. The spins interact with the Hamiltonian 
\begin{equation}
\mathcal{H}(\{S_i\}) = \frac{\xi}{2 x_T}\sum_{i,j=1}^M (S_i+q_i) (S_j+q_j) e^{-|i-j|/\xi} \equiv \frac{1}{2x_T} (S+q)^T C (S+q).
\label{hamiltonian}
\end{equation}
The system is in contact with a particle reservoir at concentration $c_\out$. Here the parameters $\xi$ and $x_T$ are dimensionless, expressed in number of lattice sites. 

The grand partition function is given by
\begin{equation}
\Xi = \sum_{S_1,\dots,S_M} z^{\sum_i |S_i|} e^{-\frac{1}{2 x_T} (S+q)^T C (S+q)},
\end{equation}
with $z = c_\out \pi R^2 L/M$ the fugacity. The matrix $C$ can be analytically inverted: 
\begin{equation}
C^{-1} = \frac{1}{2\xi \sinh (1/\xi)} \cdot 
\left(
\begin{array}{ccccccc}
e^{1/\xi} & -1 & 0 & 0 & \dots & 0 & 0 \\
-1 & 2 \cosh(1/\xi) & -1 & 0 &\dots & 0 & 0 \\
\vdots & \ddots & \ddots & \ddots &  &\vdots &\vdots \\
\vdots &  & \ddots & \ddots & \ddots &\vdots &\vdots \\
\vdots &  &  & \ddots & \ddots &\ddots &\vdots \\
0&0 &\dots &0 & -1& 2\cosh(1/\xi)& -1\\
0& 0&\dots &\dots & 0& -1& e^{1/\xi} \\
\end{array}
\right).
\end{equation}
Hence we can carry out a Hubbard-Stratonovich transformation, that is rewrite the partition function as a gaussian integral, introducing the integration variable $\varphi$: 
\begin{equation}
\Xi = \sqrt{\frac{x_T^M}{(2\pi)^M \mathrm{det} (C) }}\cdot \sum_{S_1,\dots,S_M} z^{\sum_i |S_i|} \int \d \varphi e^{-\frac{x_T}{2} \varphi^T C^{-1} \varphi + i(S+q)^T \varphi},
\end{equation}
with $\mathrm{det}(C) = \frac{e^{1/\xi}}{2 \sinh (1/\xi)}\cdot \left [ \xi (1-e^{-2/\xi}) \right]^M$. After performing the sum over the spins, which is now decoupled, we obtain
\begin{equation}
\begin{split}
\Xi & =  \sqrt{\frac{x_T^M}{(2\pi)^M \mathrm{det} (C) }}\cdot \int \d \varphi_1 \dots \d \varphi_M  \prod_{j=1}^M (1 + 2z \cos \varphi_j ) \prod_{j=1}^M e^{iq_j \varphi_j} \dots \\
& \dots \exp \left(-\frac{x_T}{4 \xi \sinh(1/\xi)}\left[\sum_{j=1}^{M-1} (\varphi_{j+1}-\varphi_j)^2+ 2(\cosh(1/\xi)-1)\sum_{j=2}^{M-1} \varphi_j^2 + (e^{1/\xi} -1)(\varphi_1^2 + \varphi_M^2)    \right]\right).
\end{split}
\label{HS}
\end{equation}
We now take a continuum limit of the lattice model. We call $a$ the physical lattice spacing and let $\tilde \xi = a \xi$, $\tilde x_T= a x_T$ and $\tilde z = M z / L$. We then let $a \to 0$ and $M \to \infty$ while keeping the physical length of the system $L = a M$ constant. We then drop the tilde sign to lighten the notation and obtain 
\begin{equation}
\Xi = \int \d \varphi(0)  e^{-x_T \varphi(0)^2/4\xi} \int [ \d \varphi ] e^{-S[\varphi]} \int \d \varphi(L)e^{-x_T \varphi(L)^2/4\xi}
\label{main_pf}
\end{equation}
with 
\begin{equation}
S[\varphi] = \int_0^L \d x  \left[\frac{x_T}{4} \left(\frac{\d \varphi}{\d x} \right)^2 + \frac{x_T}{4 \xi^2} \varphi(x)^2 - i q(x)\varphi(x)- 2z \cos \varphi(x) \right] \equiv \int_0^L \mathcal{L}(\varphi,\dot \varphi).
\label{action}
\end{equation}
$q(x)$ is the one-dimensional density corresponding to the surface charge, and $z \equiv \pi R^2 c_\out$. At this point $\xi$ and $x_T$ have the dimension of length. The path integral measure is defined as 
\begin{equation}
[\d\varphi] = \lim\limits_{\substack{a\to 0 \\ M \to \infty \\ L = aM}} \left[ \prod_{j=1}^M \sqrt{\frac{x_T}{4\pi a}} \d \varphi_j \right].
\end{equation}
We now define the propagator $P(\varphi,x|\varphi_0,0)$, or simply $P(\varphi,x)$, as 
\begin{equation}
P(\varphi,x) = \int \d \varphi(x) \delta(\varphi(x)-\varphi) \int [\d\varphi] e^{-\int_0^x \mathcal{L}(\varphi,\dot \varphi)} \int \d \varphi(0) \delta(\varphi(0)-\varphi_0).
\end{equation}
Considering an infinitesimal displacement $\Delta x$, 
\begin{equation}
\begin{split}
P(\varphi,x) = \sqrt{\frac{x_T}{4\pi \Delta x}} \int \d (\Delta \varphi) &P(\varphi - \Delta \varphi, x- \Delta x) \dots\\
&\dots \exp \left(-\int_{x-\Delta x}^x \d x'  \left[\frac{x_T}{4} \left(\frac{\Delta \varphi}{\Delta x} \right)^2 + \frac{x_T}{4 \xi^2} \varphi^2 - i q(x)\varphi- 2z \cos \varphi \right] \right).
\end{split}
\end{equation}
Expanding the propagator as $P(\varphi-\Delta \varphi, x - \Delta x) = P(\varphi,x) - \Delta x \partial P/\partial x - \Delta \varphi \partial P / \partial \varphi + (1/2) (\Delta \varphi^2) \partial^2P/\partial \varphi^2$, and carrying out the gaussian integrals, we obtain
\begin{equation}
\begin{split}
P(\varphi,x) = &\left(P(\varphi,x) - \Delta x \frac{\partial P}{\partial x} + O(\Delta x^2) \right) \left( 1 - \Delta x \left[\frac{x_T}{4\xi^2}\varphi^2 - i q(x)\varphi - 2z \cos \varphi \right] + O(\Delta x^2) \right) \\
& + \frac{\Delta x}{x_T} \frac{\partial^2 P}{\partial x^2}(1+ O(\Delta x)).
\end{split}
\end{equation}
$P(\varphi,x)$ thus solves the partial differential equation 
\begin{equation}
\frac{\partial P}{\partial x} = \frac{1}{x_T} \frac{\partial^2 P}{\partial \varphi^2} + \left(iq\varphi - \frac{x_T}{4\xi^2} \varphi^2 + 2 z \cos \varphi \right) P, 
\label{schrod}
\end{equation}
with initial condition $P(\varphi,0) = \delta(\varphi-\varphi_0)$, which is the equivalent of a Schr\"odinger equation for the path integral representation~\eqref{main_pf}. The partition function can thus be computed as 
\begin{equation}
\Xi = \int \d \varphi(L) e^{-x_T \varphi^2/4\xi} P(\varphi,L|f_0),
\label{pf}
\end{equation}
where $P(\varphi,L|f_0)$ is the solution of~\eqref{schrod} with initial condition $P(\varphi,0) = f_0 (\varphi) \equiv e^{-x_T \varphi^2/4\xi}$.

\subsection{Transfer operator}

We introduce the Fourier transform of $P$ with respect to $\varphi$: 
\begin{equation}
\tilde P (k,x) = \frac{1}{\sqrt{2\pi}} \int \d \varphi e^{-ik\varphi} P(\varphi,x).
\end{equation}
Then $\tilde P(k,x)$ satisfies
\begin{equation}
\frac{\partial \tilde P}{\partial x} = - \frac{k^2}{x_T} \tilde P - q \frac{\partial \tilde P}{\partial k} + \frac{x_T}{4 \xi^2}\frac{\partial^2\tilde P}{\partial k^2} + z \left[\tilde P(k+1,x)+\tilde P(k-1,x) \right].
\label{kspace}
\end{equation}
%
From now on, we restrict ourselves to an uncharged channel ($q = 0$). We then define the operator $\mathcal{T}$ such that 
\begin{equation}
[\mathcal{T}(\tilde P)] (k) = - \frac{k^2}{x_T} \tilde P + \frac{x_T}{4 \xi^2}\frac{\partial^2\tilde P}{\partial k^2} + z \left[\tilde P(k+1, x)+\tilde P(k-1, x) \right],
\label{transfer}
\end{equation}
which plays the role of a functional transfer matrix. Recalling eq.~\eqref{pf}, the partition function then reads
\begin{equation}
\Xi = \langle f_0 | e^{ L\mathcal{ T}} | f_0 \rangle 
\end{equation}
with $f_0(k) = e^{-\xi k^2/x_T}$ and $\langle f(k) | g(k) \rangle \equiv \int \d k f^*(k) g(k)$. 

Now, in the limit $L \to \infty$, we may consider the largest eigenvalue $\lambda$ of the operator $\mathcal{T}$, and the associated eigenfunction $\chi$: 
\begin{equation}
[\mathcal{T} ( \chi)] (k) = \lambda \chi (k).
\end{equation}
Then, up to an exponentially small correction,
\begin{equation}
\Xi = |\langle f_0 | \chi \rangle |^2  \langle \chi | \chi \rangle e^{\lambda  L}.
\label{pcharge}
\end{equation}

\subsection{Ion concentration}

Our aim is to compute the salt concentration $c_\tin$ in the nanoscale channel given a salt concentration $c_\out$ in the reservoir. At the level of the lattice model, the probability to find, say, a positive ion at position $k$, can be computed by replacing a factor $(1+2z \cos \varphi_k) $ by $z e^{i\varphi_k}$ in Eq.~\eqref{HS}. In the continuum limit, we obtain the positive (negative) ion linear density at position $x$ by inserting the operator $z e^{i\varphi}$ ($z e^{-i\varphi}$) at position $x$:
\begin{equation}
\pi R^2 \langle c_\tin^{\pm}(x) \rangle = \frac{1}{\Xi}\int \d \varphi(0)\d\varphi(x) \d \varphi(L)  e^{-x_T \varphi(0)^2/4\xi} P(\varphi(x),x|\varphi(0),0) z e^{\pm i \varphi(x)} P(\varphi(L),L|\varphi(x),x) e^{-x_T \varphi(L)^2/4\xi},
\end{equation}
Upon Fourier-transformation, the insertion of $e^{i\varphi}$ amounts to a shift by unity. Introducing the operator, 
\begin{equation}
S_Q : f \mapsto (g:k\mapsto f(k-Q)), 
\end{equation}
the concentrations are given by
\begin{equation}
\langle c_\tin^{\pm}(x) \rangle = \frac{z}{\pi R^2} \frac{ \langle f_0 | e^{ x \mathcal{ T}} S_{\pm 1} e^{( L -  x) \mathcal{ T}} | f_0 \rangle }{\Xi} = c_\out \frac{ \langle f_0 | e^{ x \mathcal{ T}} S_{\pm 1} e^{( L -  x) \mathcal{ T}} | f_0 \rangle }{\Xi},
\end{equation}
since $z = c_\out \pi R^2$. In the thermodynamic limit, and using Eq.~\eqref{pcharge} for the partition function, we obtain 
\begin{equation}
\langle c_\tin^{\pm} \rangle = c_\out \frac{\langle \chi(k) | \chi(k\mp1)\rangle}{\langle \chi(k) | \chi(k)\rangle}.
\label{cin_exact}
\end{equation}
Eq.~\eqref{cin_exact} is the main result of our exact computation. In practice, the function $\chi(k)$ is determined numerically, by finite-difference integration of Eq.~\eqref{kspace}. 

\end{widetext}

\section{Physics of ion filling} 

\subsection{Debye-H\"uckel solution}

\begin{figure*}
	\centering
	\includegraphics[width=\linewidth]{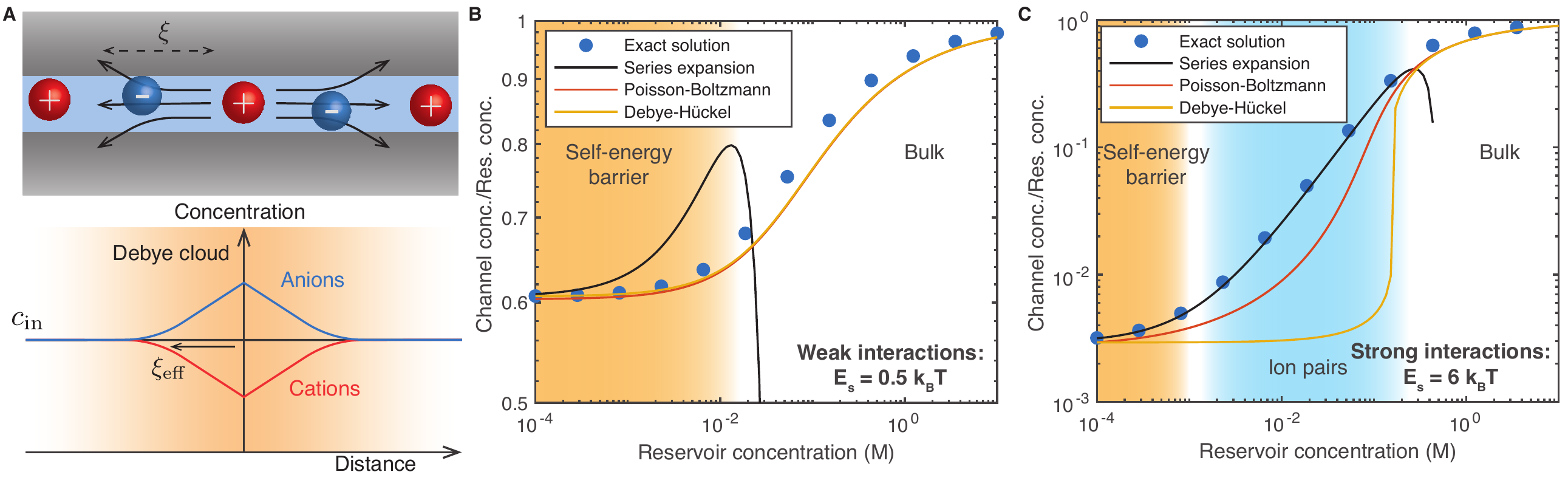}
	\caption{\textbf{Comparing mean-field approximations with the exact Coulomb gas solution.} \textbf{A}. Schematic description of the mean-field approaches. The chemical potential of confined ions is determined by solving the (linear or nonlinear) Poisson-Boltzmann equation around a given ion, interacting with an oppositely charged Debye cloud. \textbf{B}. Dependence of the channel salt concentration $c_\tin$ on the reservoir salt concentration $c_\out$, in a weakly-interacting case ($R = 1\,\si{nm}$, $\xi = 7\,\si{nm}$, $x_T = 7\,\si{nm}$, $E_\textrm{s} = 0.5 \, \kB T$). We plot four different predictions for the ratio $c_\tin/c_\out$: the exact field-theoretical solution (Eq.~\eqref{cin_exact}, blue circles), its low concentration expansion (Eq.~\eqref{expansion}, black line), the mean-field predictions from solving the full Poisson-Boltzmann equation (Eq.~\eqref{cin_mf_nl}, orange curve) or from its Debye-Hückel linearization (Eq.~\eqref{cin_mf}, yellow line). The two mean-field predictions are indistinguishable. In all cases, the naive estimate $c_\tin = c_\out$ is recovered for high enough concentrations. In the dilute limit, the concentration inside the channel is well approximated by the Arrhenius scaling $c_\tin = c_\out e^{- E_\textrm{s}/\kB T}$. \textbf{C}. Dependence of the channel salt concentration $c_\tin$ on the reservoir salt concentration $c_\out$, in a strongly-interacting case ($R = 1\,\si{nm}$, $\xi = 7\,\si{nm}$, $x_T = 0.6\,\si{nm}$, $E_\textrm{s} = 6 \, \kB T$). The color code is the same as in \textbf{B}. Here, the mean-field predictions strongly deviate from the exact solution, with the Debye-Hückel model predicting an abrupt filling transition. This discrepancy is due to the formation of Bjerrum pairs at intermediate concentrations, as evidenced by the scaling $c_\tin \propto c_\out^2$ in the exact solution.}
\end{figure*}

We now go back to the ion filling problem (Fig.~1A) and present first a one-dimensional mean-field solution. Typically, the mean-field solution of an electrolyte problem is obtained by solving the Poisson-Boltzmann equation~\cite{Andelman1995,Herrero2021}. For the conventional Poisson-Boltzmann equation to apply, we would need to consider the full three-dimensional geometry of our problem, and the effective interaction of Eq.~\eqref{potential_xT} would be introduced implicitly through the boundary conditions at the channel walls~\cite{Dresner1974}. In order to obtain a mean-field solution directly in the 1D geometry, we need to introduce a modified Poisson's equation for the electrostatic potential $\Phi$ whose Green's function coincides with Eq.~\eqref{potential_xT}: 
\begin{equation}
\left( \frac{\d^2}{\d x^2} - \frac{1}{\xi^2} \right) \phi = - 2\pi R^2 \frac{c_+ - c_-}{x_T}, 
\end{equation}
with $\phi \equiv e \Phi / k_{\rm B } T$ the dimensionless potential. Imposing that the ions follow a Boltzmann distribution ($c_{\pm} = c_\tin e^{\mp \phi}$, where $c_\tin$ is understood as the average concentration inside the channel), we obtain the analogue of the Poisson-Boltzmann equation in our 1D geometry: 
\begin{equation}
\left( \frac{\d^2}{\d x^2} - \frac{1}{\xi^2} \right) \phi = 2 \pi R^2 \frac{c_\tin}{x_T} \mathrm{sinh} \, \phi. 
\label{PB}
\end{equation}
In order to proceed analytically, we make a Debye-H\"uckel-type approximation and linearize Eq.~\eqref{PB} with respect to $\phi$. Then, the potential around an ion placed in the channel at $x = 0$ is given by 
\begin{equation}
\phi(x) = \frac{\xi_{\rm eff}}{x_T} e^{-|x|/\xi_{\rm eff}}, 
\end{equation}
with 
\begin{equation}
\xi_{\rm eff}^2 = \frac{\xi^2}{1+4 \pi R^2 c_\tin \xi^2/x_T}. 
\label{xieff}
\end{equation}
The chemical potential inside the channel is the sum of an ideal gas entropic part and of an excess part due to interactions: 
\begin{equation}
\mu_\tin = \mu_{\rm ent} + \mu_{\rm ex},
\label{muDH}
\end{equation}
with 
\begin{equation}
\mu_{\rm ent} = k_{\rm B } T \log c_\out \Lambda^3,
\label{muent}
\end{equation}
$\Lambda$ being the De Broglie thermal wavelength of the ions. $\mu_{\rm ex}$ can be obtained via a Debye charging process~\cite{Frenkel}: 
\begin{equation}
\frac{\mu_{\rm ex}}{\kB T} = \int_0^1 \phi_{\lambda}(0) \d \lambda, ~ \phi_{\lambda}(0) = \frac{\lambda \xi / x_T}{\sqrt{1+ 4 \lambda \pi R^2 c_\tin \xi^2/ x_T}}.
\label{muex}
\end{equation}
We determine $c_\tin$ by imposing equality of the chemical potentials between the channel and the reservoir:
\begin{equation}
\mu_\out = \kB  T \log c_{\rm out } \Lambda^3 = \mu_\tin, 
\end{equation}
which yields
\begin{equation}
	c_\tin = c_\out e^{- \mu_\ex/\kB T}.
	\label{chemicaleq}
\end{equation}
Evaluating analytically the integral in Eq.~\eqref{muex}, we obtain an implicit equation for $c_\tin$. With the notation ${\hat c_\tin} \equiv \pi R^2 c_\tin$, 
\begin{equation}
\begin{split}
c_\tin = c_\out \exp &\left( -\frac{\xi}{2 x_T}\times \frac{x_T^2}{6 \xi^2 {\hat c_\tin} ^2 \xi^2}\left[1 - \frac{3}{2} (1+ 4 {\hat c_\tin} \xi^2/x_T)^{1/2} \right. \right. \\  & \left. \left. + \frac{1}{2} (1+ 4 {\hat c_\tin}  \xi^2/x_T)^{3/2}   \right] \right). 
\end{split}
\label{cin_mf}
\end{equation} 
In Fig.~2\textbf{B} and \textbf{C}, we plot the ratio $c_\tin/ c_\out$ as a function of $c_\out$, as obtained by numerically solving Eq.~\eqref{cin_mf}. We fix $\xi = 7~\rm nm$ (which corresponds to a channel with $R \approx 1~\rm nm$ and strong dielectric contrast), and vary $x_T$ to set the ionic interaction strength. The interaction strength may be quantified through the self-energy barrier, $E_{\rm s} = \kB  T \times \xi / (2 x_T)$. The limiting behavior of $c_\tin/ c_\out$ may be understood directly from Eq.~\eqref{cin_mf}. When $c_\tin$ is small, Eq.~\eqref{cin_mf} reduces to \PR{the Arrhenius scaling} $c_\tin = c_\out e^{-E_{\rm s} / k_{\rm B }T}$: this results typically holds for biological ion channels which may contain either 0 or 1 ion at any given time, and the effect of inter-ionic interactions is negligible. When $c_\tin$ is large, we recover $c_\tin = c_\out$. Indeed, the excess term in the chemical potential vanishes at high concentrations, which is then dominated by the entropic term. The fact that $\mu_{\rm ex} \to 0$ as $c_\tin \to \infty$ is non-trivial: it can be seen, physically, as resulting from the Coulomb potential of each ion being perfectly screened by the other ions. At small values of $E_{\rm s}$, Eq.~\eqref{cin_mf} has a single solution for all values of $c_\out$, which interpolates smoothly between the two limiting regimes. However, for $E_{\rm s} \gtrsim 5 \kB  T$, it has three solutions in a certain range of $c_\out$, pointing to a pseudo-first-order phase transition between a low-concentration and a high-concentration phase, similar to the one predicted by Dresner~\cite{Dresner1974} and Palmeri \emph{et al.}~\cite{Buyukdagli2010}. The transition occurs at ${\hat c_\tin} \sim x_T / \xi^2$: as per Eq.~\eqref{xieff}, this corresponds to the concentration where the effect of the screening cloud on an ion's Coulomb potential becomes significant. 

\subsection{Full Poisson-Boltzmann solution}

The physical content of the mean-field solution presented above is similar to the one of Dresner, based on a linearized Poisson-Boltzmann equation~\cite{Dresner1974}. The difference in geometry, and the fact that he foregoes the use of the Debye charging process, do not seem to play a significant qualitative role. The solution of Palmeri \emph{et al.}~\cite{Buyukdagli2010} takes ionic correlations into account to some extent, yet it still involves a Debye-H\"uckel-type linear equation for the mean-field interaction potential between the ions. 

One may ask whether the same phenomenology persists if one does not linearize the Poisson-Boltzmann equation. \PR{The full Poisson-Boltzmann equation cannot be solved analytically, but supports the following integral form:
\begin{equation}
	\left( \frac{\d \phi}{\d x} \right)^2 - \frac{1}{\xi^2}\phi^2 = 4 \pi R^2 \frac{c_\tin}{x_T} \left(\cosh \phi - 1 \right),
	\label{pb_integral}
\end{equation}
where we have used the fact that $\phi$ should vanish at $x \to \infty$. For $x \to 0$, the solution of Eq.~\eqref{pb_integral} should reduce to the unscreened potential in Eq.~\eqref{potential_xT} up to an additive constant, so that
\begin{equation}
	\frac{1}{x_T^2} - \frac{1}{\xi^2}\phi^2(0) = 4 \pi R^2 \frac{c_\tin}{x_T} \left(\cosh \phi(0) - 1 \right).
\end{equation}

Once again, one may express the excess chemical potential of the confined ions through a Debye charging process:
\begin{equation}
	\begin{split}
	\frac{\mu_{\rm ex}}{\kB T} &= \int_0^1 \phi_{\lambda}(0) \d \lambda,\\ \frac{\lambda^2}{x_T^2} - \frac{1}{\xi^2}\phi_\lambda^2(0) &= 4 \pi R^2 \frac{\lambda c_\tin}{x_T} \left(\cosh \phi_\lambda(0) - 1 \right).
	\end{split}
	\label{muex_nl}
\end{equation}
This result is the analogue of Eq.~\eqref{muex}, with $\phi_\lambda(0)$ now being the solution of an implicit non-linear equation, so that $\mu_\ex$ must be determined numerically. As before, the concentration inside the channel is then given by:
\begin{equation}
	c_\tin = c_\out e^{- \mu_\ex/\kB T}.
	\label{cin_mf_nl}
\end{equation}
}

The prediction of the full Poisson-Boltzmann equation is shown in Fig.~2\textbf{B} and \textbf{C}: we find $c_\tin$ to be a smooth function of $c_\out$ for all values of parameters, in contrast to the linearized solution. We may not, however, unambiguously conclude that the filling transition is an artifact of linearization, since the non-linear solution still involves a mean-field approximation and is not guaranteed to yield the correct result. 

Interestingly, the ``physically-motivated" mean-field solution in Eq.~\eqref{PB} differs from the mean-field limit of our exact solution. It is obtained by taking the saddle-point approximation in the path-integral expression of the partition function (Eq.~\eqref{main_pf}). The Euler-Lagrange equation for the minimizer $\varphi(x)$ of the action $S[\varphi]$ in Eq.~\eqref{action} is, upon identifying $\phi = -i \varphi$, 
\begin{equation}
\left( \frac{\d^2}{\d x^2} - \frac{1}{\xi^2} \right) \phi = 2 \pi R^2 \frac{c_\out}{x_T} \mathrm{sinh} \, \phi. 
\label{PB2}
\end{equation}
This is Eq.~\eqref{PB} with $c_\tin$ replaced with $c_\out$, and corresponds to a first order treatment of interactions. Indeed, if the ions are non-interacting, $c_\tin = c_\out$. By solving the mean-field equation, we determine how the ions' chemical potential is affected by Debye screening, which then results in value of $c_\tin$ that is different from $c_\out$. Within a straightforward interaction expansion procedure, one should determine the effect of screening assuming the zeroth order value for the ion concentration inside the channel, which is $c_\out$: this corresponds to Eq.~\eqref{PB2}. Eq.~\eqref{PB} contains an additional self-consistency condition, as it assumes the actual value $c_\tin$ for the ion concentration, which is not known until Eq.~\eqref{PB} is solved. One may draw a loose condensed matter physics analogy, where Eq.~\eqref{PB2} resembles the Born approximation for impurity scattering, while Eq.~\eqref{PB} is analogous to the self-consistent Born approximation.~\cite{Bruus}

\subsection{Exact solution}

We now turn to the exact solution obtained in Sec. II to unambiguously solve the ion filling problem. We determine $c_\tin$ according to Eq.~\eqref{cin_exact}: 
\begin{equation}
\langle c_\tin^{\pm} \rangle = c_\out \frac{\langle \chi(k) | \chi(k\mp1)\rangle}{\langle \chi(k) | \chi(k)\rangle}, 
\end{equation}
where $\chi(k)$ is the highest eigenfunction of the transfer operator in Eq.~\eqref{transfer}, determined in practice by numerical integration. The exact results for $c_\tin$, with the same parameter values as for the mean-field solution, are shown in Fig.~2 \textbf{B} and \textbf{C}. When interactions are weak (small values of $E_s$, Fig.~2\textbf{B}), the exact and mean-field solutions are in good agreement. \PR{Notably, all solutions smoothly interpolate between the bulk scaling $c_\tin = c_\out$ at high concentration, and the Arrhenius scaling $c_\tin = c_\out e^{- E_\textrm{s}/\kB T}$ at low concentration.} Conversely, in the strongly-interacting case (large $E_s$, Fig.~2\textbf{C}), the exact result yields a much larger ion concentration that the mean-field solutions for intermediate values of $c_\out$. In this intermediate regime, $c_\tin$ remains a smooth function of $c_\out$, and obeys the scaling $c_\tin \propto c_\out^2$.

 Such a scaling is the signature of the formation of tightly bound Bjerrum pairs of positive and negative ions -- strongly-correlated configurations that are not taken into account by mean-field solutions. Indeed, let us assume that the channel contains an ideal gas of ion pairs at concentration $c_\tin$. We further assume that in a pair, the distance between the two ions is uniformly distributed in the interval $[-x_T/2,x_T/2]$, and the binding energy of a pair is $\kB  T \xi/x_T = 2E_s$. Then, the grand partition function reads
\begin{align}
\Xi &= \sum_N (ze^{-\beta E_s})^{2N} \frac{1}{N!} \prod_{i=1}^{N} L\int_{-x_T/2}^{x_T/2} \d x \, e^{2\beta E_s} \\
& = \sum_N \frac{( z^2 L x_T )^N}{N!} = e^{ z^2 Lx_T},
\end{align}
\PR{where we recall that $z = \pi R^2 c_\out$ and $\beta \equiv 1/(k_{\rm B} T)$.} Using that 
\begin{equation}
\pi R^2 c_\tin =  \frac{1}{L} \frac{\partial \log \Xi}{\partial (\beta \mu)} = \frac{z}{L} \frac{\partial \log \Xi}{\partial z}, 
\end{equation}
we obtain 
\begin{equation}
c_\tin = \frac{2 z^2 x_T}{\pi R^2} = 2 \pi R^2 x_T c_\out^2.
\label{cin_scaling}
\end{equation}
We recover indeed the quadratic scaling. 

We may check that the prefactor in Eq.~\eqref{cin_scaling} is the correct one by evaluating analytically the expression in Eq.~\eqref{cin_exact} in the low concentration limit $z_T \equiv zx_T \ll 1$. An analytical expansion of the function $\chi(k)$ in powers of $z_T$ was derived in ref.~\cite{Kavokine2019}. Substituting it into Eq.~\eqref{cin_exact}, we obtain 
\begin{equation}
\begin{split}
\pi R^2 c_\tin = z(e^{-\beta E_s}+2 z_T-\frac{13}{2} z_T^2 e^{-\beta E_s}  \\ -7z_T^3+ O(z_T^4) + O(e^{-2\beta E_s}) ).
\end{split}
\label{expansion}
\end{equation}
The first term in the expansion corresponds to $c_\tin = c_\out e^{-\beta E_s}$. At the lowest salt concentrations, forming Bjerrum pairs is too entropically unfavorable, and the concentration inside the channel is controlled by the self-energy barrier. However, as the salt concentration increases, there is no abrupt transition to a highly-screened concentrated phase inside the channel; instead, the channel is progressively filled by Bjerrum pairs. This corresponds to the quadratic term in the expansion, with the prefactor agreeing indeed with Eq.~\eqref{cin_scaling}.\footnote{This justifies \emph{a posteriori} our choice of $[-x_T/2, x_T/2]$ as the interval in which a paired-up ion is allowed to move.} The expansion in Eq.~\eqref{expansion} reproduces quite well the low-concentration behavior of the exact solution as shown in Fig.~2\textbf{B} and \textbf{C}. However, it fails at high concentrations, where it does not recover $c_\tin = c_\out$. 

Our exact analysis of the ion statistics in a nanoscale channel has revealed that Bjerrum pairs are a crucial ingredient of the filling process. We now develop a modified mean-field theory that accounts the presence of Bjerrum pairs and compare it to the exact solution. 

\section{Pair-enhanced mean-field theory}

\subsection{Debye-Hückel-Bjerrum theory}

\begin{figure*}
	\centering
	\includegraphics[width=\linewidth]{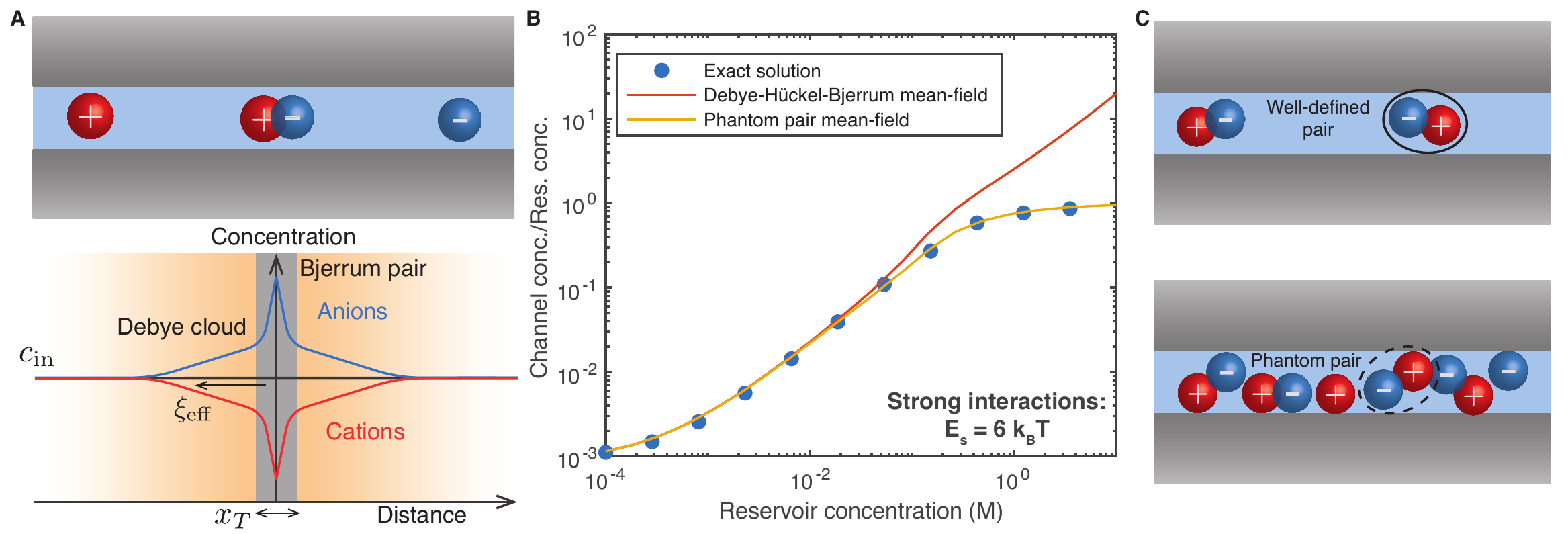}
	\caption{\textbf{Pair-enhanced mean-field theory.} \textbf{A}. Treatment of ion pairing in mean-field approaches. Top panel: Mean-field theories inevitably underestimate ion-ion correlations. To circumvent this problem, two ions that are distant by less than $x_T$ are considered to form an ion pair, which is treated as a separate chemical species. Bottom panel: schematic representation of ion distribution around a fixed positive ion. The distribution is very peaked close to the central ion, due to the formation of an ion pair, and then relaxes smoothly to the mean value $c_\tin$. \textbf{B}. Evolution of channel concentration $c_\tin$ as function of reservoir concentration $c_\out$, in a strongly-interacting cacse ($R = 1\,\si{nm}$, $\xi = 7\,\si{nm}$, $x_T = 0.6\,\si{nm}$, $E_\textrm{s} = 6 \, \kB T$). We plot the ratio $c_\tin/c_\out$ obtained from three different models taking Bjerrum pairs into account: the exact field-theoretical solution (Eq.~\eqref{cin_exact}, blue circles), the Debye-Hückel-Bjerrum mean-field theory (Eq.~\eqref{cin_DHBj}, red line) and our modified mean-field theory based on the notion of phantom pairs (Eq.~\eqref{cin_mf3}, orange line), which reproduces the exact solution quantitatively for all values of parameters. At high concentration, the Debye-Hückel-Bjerrum prediction fails due to the uncontrolled proliferation of Bjerrum pairs. \textbf{C}. Formation of phantom pairs inside the nanochannel. At low concentration (top panel), pairs are well-separated and ions forming a pair are tightly bound to each other. At high concentration (bottom panel), ionic interactions are weakened as a result of Debye screening, and two quasi-non-interacting ions may find themselves within a distance $x_T$ of each other without actually binding: this is a phantom pair.}
\end{figure*}

The traditional mean-field treatment of electrolytes is incapable of taking Bjerrum pairs into account, as it naturally neglects any strong ion-ion correlations -- pairing being a fundamentally discrete phenomenon. An idea proposed by Bjerrum to amend the Debye-Hückel theory was to introduce ion pairs as a separate species encapsulating all ``strong'' ion-ion correlations \cite{Levin2002}. More precisely, any two oppositely charged ions that are closer than some minimum distance can be considered as a single neutral entity -- a Bjerrum pair. The remaining ``free'' ions should then only experience weak interactions with each other, and can be treated at the mean-field level. Importantly, this last remark justifies the Debye-Hückel linearization, as all non-linear effects are assumed to be hidden in the definition of ion pairs.

As before, we consider that pairs behave like particles of an ideal gas, and that their maximum extension is given by $x_T$. Defining $c_\tin^\p$ the concentration pairs inside the channel, the chemical potential of pairs is given by:
\begin{equation}
	\mu_\tin^\p = \kB  T \log \frac{c_\tin^\p \Lambda^6}{2\pi x_T R^2},
\end{equation}
where the geometrical factor inside the logarithm accounts for the internal degrees of freedom of a pair. The chemical potential only has an entropic term, because the binding energy of the pair exactly compensates the self-energy of the two separate ions. The chemical equilibrium between free ions and pairs inside the channel can be written as:
\begin{equation}
	\mu_\tin^+ + \mu_\tin^- = 2 \mu_\tin = \mu_\tin^\p,
\end{equation}
where $\mu_\tin^+$ and $\mu_\tin^-$ are the chemical potentials of cations and anions, respectively. We then obtain, using the Debye-Hückel solution for $\mu_\tin$ (equations \eqref{muDH} to \eqref{muex}):
\begin{equation}
	c_\tin^\p = 2 \pi R^2 x_T c_\out^2,
\end{equation}
which is the result obtained in the previous section. The average concentration in free ions $c_\tin^\f$ is not modified compared to the Debye-Hückel solution, and is therefore the solution of the self-consistent Eq.~\eqref{cin_mf}. One can then compute the total concentration inside the channel as $c_\tin = c_\tin^\f + c_\tin^\p$, or, explicitly
\begin{equation}
	c_\tin = c_\out e^{-\mu_{\rm ex} (c_\tin^\f)/\kB T}+ 2 \pi R^2 x_T c_\out^2.
	\label{cin_DHBj}
\end{equation}
In other words, the only impact of pairs in Bjerrum's computation is to add a quadratic term $2 \pi R^2 x_T c_\out^2$ to the Debye-Hückel result, matching with the expansion \eqref{expansion} of the exact solution up order 2 in the bulk concentration. We compare the two predictions on Fig.~3\textbf{B}. The Debye-Hückel-Bjerrum solution is found to match the exact one quite well at low and intermediate concentrations. This result is, however, unphysical for $c_\out \gtrsim 1/\pi R^2 x_T$: $c_\tin$ is found to grow much faster than the bulk concentration. One solution would be to consider higher-order terms in the mean-field treatment through the inclusion of triplets, quadruplets, etc. of ions, and all possible interactions between these entities. Truncating the sum at any finite order, however, would not yield a solution valid in the entire range of concentrations, nor is it guaranteed to converge to the exact solution. This approach is also unsatisfactory as it would not yield a closed-form expression for $c_\tin$ and would not allow for qualitative understanding of the underlying physics.

Instead, we develop a different method that, through physics-driven arguments, prevents the divergence of $c_\tin$ at high bulk concentrations and reproduces quantitatively the exact solution.

\subsection{Phantom pairs}

Eq.~\eqref{cin_DHBj} overestimates the number of Bjerrum pairs in the channel because it fails to account for the presence of Bjerrum pairs in the reservoir. The electrolyte in the reservoir is treated as an ideal gas : the ions are non-interacting and they cannot form actual tightly-bound pairs. Nevertheless, we have defined any two oppositely charged ions that find themselves in a cylinder of radius $R$ and length $x_T$ to be a separate chemical species. Such configurations may arise in the reservoir simply out of statistical chance: we dub them \emph{phantom pairs}. For our mean-field theory to be consistent, these phantom pairs need to be taken into account. 


Let $c_\out^\p$ be the concentration of phantom pairs in the reservoir. The chemical equilibrium between phantom pairs and free ions imposes
\begin{equation}
	c_\out^\p = 2 \pi R^2 x_T (c^\f_\out)^2.
\end{equation}
In addition, one has $c_\out^\f + c_\out^\p = c_\out$, since an ion must either be free or part of a pair. Imposing this condition yields:
\begin{equation}
	c_\out^\f = \frac{\sqrt{1 + 8 \pi c_\out x_T R^2}-1}{4 x_T \pi R^2}.
	\label{cfout}
\end{equation}
We use this result to control the proliferation of pairs in the channel: we now equilibrate the free ions inside the nanochannel with only the free ions in the reservoir:
\begin{equation}
		c_\tin^\f  = c_\out^\f e^{-\mu_\ex(c_\tin^\f) / \kB T},
		\label{chemicaleq_phantom}
\end{equation}
which corresponds to Eq.~\eqref{chemicaleq} with $c_\out$ replaced by $c^f_\out$. Eq.~\eqref{chemicaleq_phantom} is again a self-consistent equation, this time on the concentration of free ions $c^\f_\tin$, that must be solved numerically. Lastly, equilibrating pairs with free ions inside the channel (or, equivalently, pairs inside with pairs outside), we obtain:
\begin{equation}
	c_\tin = c_\tin^\f + 2 \pi R^2 x_T (c_\out^\f)^2,
	\label{cin_mf3}
\end{equation}
where the second term corresponds again to Bjerrum pairs. Eqs.~\eqref{cfout} to \eqref{cin_mf3} constitute the main result of our modified mean-field theory. Note that $\mu_\ex$ may be determined at the Debye-Hückel level (Eq.~\eqref{muex}), or by solving the full Poisson-Boltzmann equation (Eq.~\eqref{muex_nl}). In what follows, we will only discuss the latter, as it offers greater accuracy; however, the Debye-Hückel prediction provides reasonable results even in the case of strong interactions, and yields for a convenient analytical expression for $\mu_\ex$ as function of $c_\tin^\f$.

The prediction of our phantom pair Poisson-Boltzmann model is compared to the exact solution \eqref{cin_exact} in Fig.~3\textbf{B}. The two solutions are found to be in near perfect agreement for all values of parameters, even in strong coupling limit $E_\textrm{s} \gg \kB T$.

In the next two sections, we use our modified mean-field model to predict the conductance of a nanochannel, first in the case of a neutral channel, and then in presence of a surface charge.

\subsection{Conductance}

\begin{figure*}
	\centering
	\includegraphics[width=0.8\linewidth]{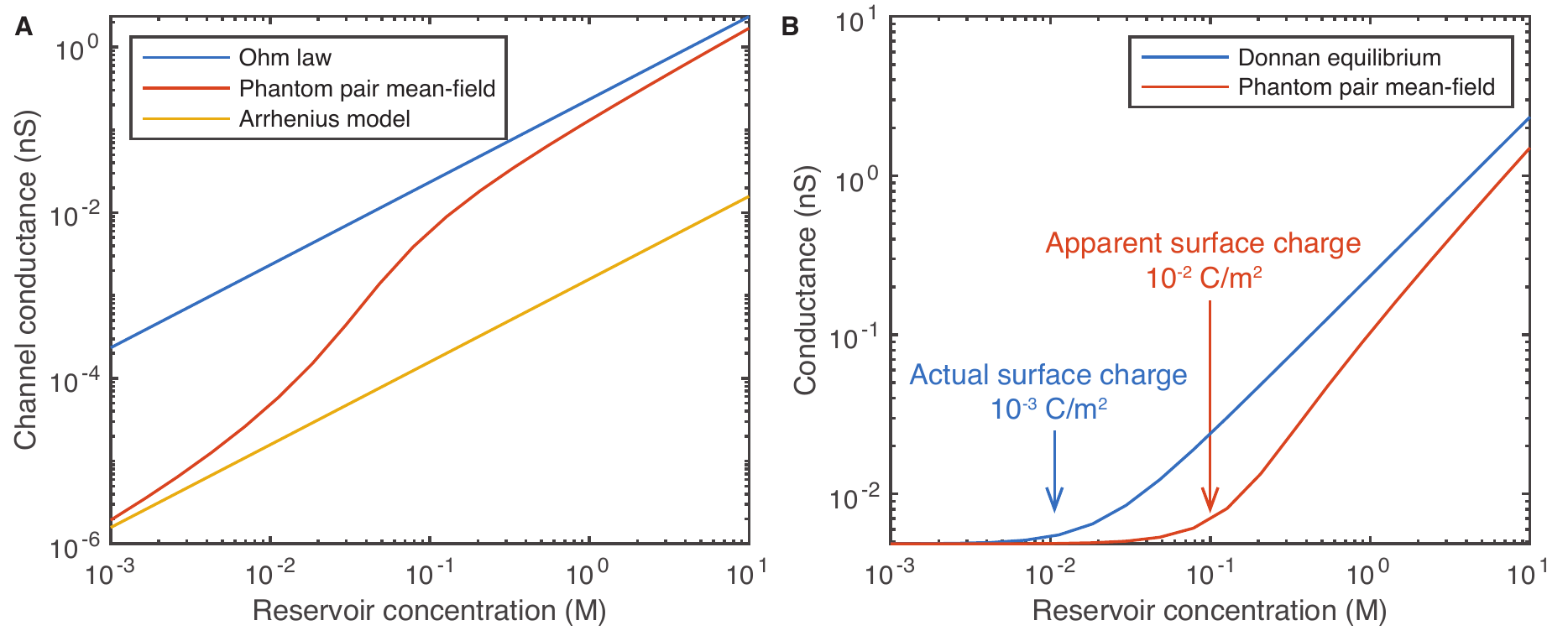}
	\caption{\textbf{Channel conductance in the pair-enhanced mean-field model.} \textbf{A}. Conductance of a nanochannel ($R = 1\,\si{nm}$, $\xi = 7\,\si{nm}$, $x_T = 0.7\,\si{nm}$, $E_\textrm{s} = 10 \, \kB T$) as function of the reservoir concentration. The red line corresponds to the prediction of the phantom pair mean-field model (Eq.~\eqref{conductance}) for $T = 300 \, \si{K}$, $D = 10^{-9} \, \si{m^2/s}$ and $L = 100 \, \si{nm}$. The Ohm's law bulk prediction ($c_\tin = c_\out$, blue line) and the Arrhenius model ($c_\tin = c_\out e^{- E_\textrm{s}/\kB T}$, yellow line) are also represented for comparison. \textbf{B}. Conductance of a nanochannel with a weak surface charge $\Sigma = 10^{-3} \, \si{C/m^2}$. We represented the predictions of the conventional Donnan equilibrium (Eq.~\eqref{donnan}, blue line) and of the phantom pair mean-field theory (equations \eqref{conductance} and \eqref{modifieddonnan}, red line). Because interaction confinement results in a lower ion concentration in the channel, the usual formula $\Sigma \sim R c^*/2$, where $c^*$ is the reservoir concentration for which conductance levels off overestimates the surface charge by one order of magnitude, as indicated on the plot.}
\end{figure*}

One strength of our modified mean-field model is that it offers insight into the physical properties of the confined system beyond the value of the ionic concentration. In particular, the decomposition of the electrolyte into free ions and bound pairs allows us to estimate the channel's conductance. Tightly bound Bjerrum pairs are electrically neutral, so that they do not contribute to the ionic current to first order in applied electric field: it would then be straightforward to assume that the channel's conductance is proportional to the concentration of free ions. However, the reasoning needs to be more subtle, since the channel, in the same way as the reservoir, may contain non-interacting phantom pairs. Indeed, we have decomposed the confined electrolyte into tightly bound pairs, that have no ionic atmosphere, and free ions that are dressed by a Debye screening cloud. As the concentration increases, the interaction  between dressed ions becomes weak, and two of them may find themselves within a distance $x_T$ without actually binding. Such a phantom pair is expected to still contribute to the conductance. The concentration of phantom pairs in the channel is obtained by imposing their chemical equilibrium with the free ions treated as an ideal gas. 
Thus, we estimate the channel's conductance as:
\begin{equation}
	G = 2\frac{e^2 D}{\kB  T} \frac{\pi R^2}{L} \left(c_\tin^\f + 2 x_T \pi R^2 (c_\tin^\f)^2\right),
	\label{conductance}
\end{equation}
where $D$ is the diffusion coefficient of ions; the second term corresponds to the contribution of phantom pairs. In Fig.~4\textbf{A}, we compare this result to the Ohm's law prediction where pairs are neglected and one assumes $c_\tin = c_\out$. Ohm's law is found to greatly overestimate the conductance at low concentration. In the dilute limit, we instead recover the Arrhenius scaling, where one assumes $c_\tin = c_\out e^{-E_\textrm{s}/\kB T}$.

Finally, we stress that Eq.~\eqref{conductance} only accounts for the electrophoresis of free ions, and is therefore only valid in the limit of weak external electric fields. Stronger voltage drops will result in the breaking of ion pairs, causing a conductivity increase in a process known as the second Wien effect. This phenomenon is described in refs. \cite{Kavokine2019,Robin2021}, and has been used to create solid-state voltage-gated nanochannels\cite{Robin2023}.

\subsection{Effect of a surface charge}

Up till now, we have restricted ourselves to channels with uncharged walls. However, in most experimentally relevant situations, the channel walls bear a surface charge density $\Sigma$, which strongly impacts nanofluidic transport. While introducing a surface charge is tedious within the exact framework, we may readily assess the effect of surface charge in the interaction confinement regime using our pair-enhanced mean-field theory. 

In the limit where the channel's radius is smaller than the Debye length, we assume that the presence of the surface charge amounts to a homogeneous Donnan potential drop $V_\D$ inside the channel, which we do not need to determine explicitly. Then, the chemical potential of ions inside the channel reads:
\begin{equation}
	\mu_\tin^\pm = \mu_\text{ex} \pm eV_\D + \kB  T \log c_\tin^\pm \Lambda^3.
\end{equation}
Note that the concentration in free anions $c_\tin^-$ and cations $c_\tin^+$ are now distinct, so that $\mu_\ex$ is defined as a function of the average free ion concentration $c_\tin^\f= (c_\tin^+ + c_\tin^-)/2$. In a channel that is sufficiently long for local electroneutrality to hold, 
\begin{equation}
	c_\tin^+ - c_\tin^- + 2 \Sigma/R = 0. 
\end{equation}
Imposing chemical equilibrium with the reservoir, we obtain a modified version of the Donnan result (Eq.~\eqref{donnan}): 
\begin{equation}
\left\{
\begin{array}{l}
c_\tin = c_\tin^\f + c_\tin^p \\
\\
c_\tin^\f = \sqrt{\left(c_\out^\f e^{- \beta \mu_\text{ex}(c_\tin^\f)}\right)^2 + \left( \frac{2 \Sigma}{R}\right)^2}, \\
\\
c_\tin^\p =  2 \pi R^2 x_T (c_\out^\f)^2,
\end{array}
\right.
\label{modifieddonnan}
\end{equation}
with $c_\out^\f$ given by Eq.~\eqref{cfout}. 

One can again obtain the channel's conductance through Eq.~\eqref{conductance}, which we compare to the Donnan / Ohm's law result  in Fig.~4\textbf{B}. Importantly, the Donnan result predicts that conductance becomes independent of concentration for $c_\out \sim 2 \Sigma/R$ (see Eq.~\eqref{donnan}). In practice, this result is commonly used to estimate experimentally the surface charge as $\Sigma \sim R c^*/2$, where $c^*$ is the reservoir concentration for which conductance levels off.  In contrast, in the interaction confinement regime, we predict that the transition occurs instead at $c_\tin^\f \sim 2\Sigma/R$ --  corresponding to a higher reservoir concentration, due to the self-energy barrier. In this case, Donnan's prediction overestimates the surface charge by typically one order of magnitude, as shown in Fig.~4\textbf{B}.

Finally, let us stress that we considered here a charge homogeneously distributed along the channel's surface. This assumption is relevant in the case of conducting wall materials, such as systems where the charge is imposed via a gating electrode connected to the channel walls. This situation, however, may be different in experimentally-available devices, where the surface charge generally consists in localized charged groups and defects on the channel walls. In this case, the physics become more involved as ions may form bound pairs with the fixed surface charges. Some of these physics have been revealed by the exact computations of Shklovskii and coworkers~\cite{Zhang2005,Zhang2006}; a technically simpler approach to these physics using our pair-enhanced mean-field theory would be possible, but extends beyond the scope of the present work.

\section{Discussion and perspectives}

We have determined the salt concentration inside a nanometric channel connected to reservoirs filled with electrolyte. In the case of a fully 1D geometry, corresponding to a nanotube of radius $R \sim 1 \si{nm}$, we developed an exact field-theoretical solution that allowed us to compute channel concentration $c_\tin$ as function of the reservoir concentration $c_\out$. This solution clears up the ambiguities of pre-existing mean-field theories, and contradicts the naive expectation $c_\tin = c_\out$. In particular, the concentration inside the nanochannel is found to be always lower than in the bulk, as the confinement of electrostatic interactions creates an energy barrier for ions to enter the channel.

Yet, we found that $c_\tin$ is in fact higher than the prediction of the mean-field Debye-Hückel theory, as ion pairing is counterbalances to some extent the energy cost of interaction confinement. Such strong ion-ion correlations cannot be directly accounted for in a mean-field theory, and the filling transition that emerges in Debye-H\"uckel theory appears to be an artefact of linearization. To overcome this issue, one can add Bjerrum pairs as a separate chemical species within the Debye-Hückel model. Carefully accounting for the statistical formation of unbound \emph{phantom pairs}, we obtain a modified mean-field theory that reproduces the result of the exact computation with nearly-perfect accuracy, and that can be extended to account for a non-zero surface charge on the channel wall.

Despite the concurring results, the two original formalisms developed in this work serve distinct purposes. The field-theoretical solution plays the role of a touchstone model, owing to its exact treatment of all many-body interactions. Modeling electrolytes is a notoriously hard problem in statistical physics, and simplified models often lack a lack a reference solution for benchmarking their approximations. This difficulty is lifted in the 1D geometry: thanks to the existence of the exact solution, we have been able to build a quantitatively precise mean-field model, adding step-by-step the qualitative ingredients necessary to reproduce the exact result. 

Moreover, the field theory formalism gives access to the entire statistics of the system, including finite-size effects which elude any mean-field treatment. The latter are expected to be relevant in many experimental situations, as a substantial amount of current works focuses on short pores, where the length of the channel is comparable to its radius. For instance, one can expect shorter channels to deviate from electroneutrality \cite{Levy2020} -- something entirely impossible in the limit of infinitely long channels. 

On the other hand, our modified mean-field formalism has the advantage of mathematical simplicity, allowing for convenient physical interpretations. The simple distinction between free ions and Bjerrum pairs can be used to straightforwardly estimate the channel's conductance. The influence of ion-ion correlations on conductivity is of particular importance as conductance measurements underpin many nanofluidic experiments. In contrast, the exact solution does not provide any such insight on transport properties, as it is limited to thermal equilibrium. 

Furthermore, the mean-field model may easily be adapted to other geometries, whereas an exact treatment is only possible in the strictly 1D case. Extensions of our results to 2D nanochannels would be of significant interest. In particular, 2D nanochannels can be made out of various materials with different electronic properties, which directly impact the confined ionic interactions~\cite{Kavokine2022}. Therefore, 2D nanochannels could serve as a platform for exploring the impact of wall metallicity on the ion filling problem.

Both our exact and mean-field solutions can be expected to fail at very high concentrations. Indeed, our work relies on a simplified picture of electrolytes, where all steric effects are discarded. We considered point-like ions with no short-distance repulsion; therefore, no effect like saturation or layering can be accounted for. Similarly, we neglected any interaction with the solvent -- for example, we did not consider the decrement in relative permittivity at high salt concentrations \cite{Levy2013}. However, since all electrostatic interactions are screened in the limit of high concentrations, such considerations should not impact the conclusions of the present work: particularly, we would still expect that $c_\tin = c_\out$ at high concentration. 

Lastly, let us briefly recall our results for the ion filling problem. In channels larger than a few nanometers, the conventional mean-field picture is valid, so that in absence of any surface charge the salt concentration inside the channel equals that of the reservoirs: $c_\tin = c_\out$. For nanometre-scale confinement and low concentrations, interaction confinement amounts to a finite energy barrier for ions to enter the channel: $c_\tin = c_\out e^{- E_{\rm s}/\kB T}$. As concentration increases, more ions are able to overcome the barrier by forming Bjerrum pairs, neutralizing the electrostatic cost of confinement, at the price of entropy: $c_\tin \propto c_\out^2$. Only at high concentrations can one recover the intuitive estimate $c_\tin = c_\out$, as intense screening cancels out all electrostatic interactions. Overall, interaction confinement has a significant impact on the properties of nanofluidic systems, and the assumption $c_\tin = c_\out$ should be questioned any time the system's size reaches the nanometre scale.

\begin{acknowledgments}
N.K. acknowledges support from a Humboldt fellowship. L.B. acknowledges funding from the EU H2020 Framework
Programme/ERC Advanced Grant agreement number 785911-Shadoks. The Flatiron Institute is a division of the Simons Foundation.
\end{acknowledgments}

\section*{Data Availability Statement}

The data that support the findings of this study are available from the corresponding author upon reasonable request.

\end{document}